\documentclass[12pt]{article}

\usepackage{graphicx}
\usepackage{amssymb,amsmath}
\usepackage{mathtools}
\usepackage{geometry}
\usepackage{array}

\begin{document}

\begin{titlepage}

\begin{flushright}
arXiv:2506.19564
\end{flushright}
\vskip 2.5cm

\begin{center}
{\Large \bf Dalitz Plot Kinematics for a\\
Lorentz-Violating Three-Body Decay}
\end{center}

\vspace{1ex}

\begin{center}
{\large Joshua O'Connor and Brett Altschul\footnote{{\tt altschul@mailbox.sc.edu}}}

\vspace{5mm}
{\sl Department of Physics and Astronomy} \\
{\sl University of South Carolina} \\
{\sl Columbia, SC 29208} \\
\end{center}

\vspace{2.5ex}

\medskip

\centerline {\bf Abstract}

\bigskip

Rates for particle interaction processes and decays will be modified in a Lorentz-violating quantum field
theory, because of changes to the particle kinematics---particularly through the modified dispersion
relations affecting the outgoing particle phase space. We outline here these changes
to the rates for three-particle decays. Considering a process with a constant scattering amplitude (not directly
modified by the Lorentz violation), we calculate leading order corrections to the kinematics for a
decay into three identical spinless particles whose propagation is affected by a $c_{\mu\nu}$-type
symmetric tensor background. We examine the angular distribution of the daughter particles and
describe the shape of the corresponding Dalitz plot outlining the kinematically allowed region, according
to two toy models for the $c_{\mu\nu}$ textures. Precision measurements of the boundaries of this region could
be used to constrain Lorentz violation coefficients for the particles involved in processes such as
$\eta\rightarrow 3\pi^{0}$. 

\bigskip

\end{titlepage}

\newpage

\section{Introduction}

Research in modern particle theories emphasizes the calculation of transition matrix elements, meaning
the dynamics.
However, the kinematics of scattering and decay processes can be just as important or in many cases
more important than the matrix elements $i\mathcal{M}$. The reason for this is 
that the kinematical and the dynamical parts of a differential rate calculation can be separated out rather
neatly. For a constant matrix element, even in the fully integrated rate the effects of kinematics and dynamics
can be calculated as independent multiplicative factors, although more generally the two are mixed in
the integration over all channels. Since the kinematic factors depend only on the behavior of the free
asymptotic incoming and outgoing particles, they do not need to be calculated anew for each process---at least
in a conventional theory with standard energy-momentum relations for all the particle species.

In the standard model, physics must be identical in each inertial frame: invariant
under Lorentz transformations. However, we relax this notion and implement particle
theory models in inertial frames where the physics is modified. These Lorentz noninvariant theories
contain inertial frames that are not all equivalent. This makes
necessary changes to specific calculation procedures used in the standard model; we
can obtain completely or slightly modified observable values or predict new physical
phenomena that can also be measured. For example, changes in the standard energy-momentum relations for
various types of quanta can affect the rates and thresholds
for particle processes.

Theories with Lorentz symmetry breaking are also of interest, since quantum theories of gravitation
may not have the same symmetries as the low-energy effective theories that we see in everyday operation.
Theories of Planck-scale physics like strings, loop quantum gravity, noncommutative spacetime structures,
spacetime foam, nontrivial spacetime topologies, and others already predict or strongly suggest Lorentz
violation~\cite{ref-kost18,ref-kost19,ref-gambini,ref-klinhamer3,ref-mocioiu,ref-amelino22,ref-klinkhamer4,
ref-carroll3,ref-alfaro,ref-klinkhamer,ref-bojowald,ref-bernadotte,ref-hossenfelder,ref-ghosh}.
The same tools used to study the standard model may also be used to analyze more general theories in which
Lorentz symmetry is weakly broken.
The general local field theory that describes Lorentz-violating modifications to
processes involving known standard model fields has been laid out in detail~\cite{ref-kost1,ref-kost2}.
This theory, which is called the standard model extension (SME), is also the general theory for
describing local CPT violation involving those known fields (although, while the SME
may be extended to include gravity~\cite{ref-kost12}, gravitational CPT violation is
a more complicated matter~\cite{ref-jackiw5,ref-guarrera,ref-alexander1});
the reason is that a local, stable, unitary, CPT-violating quantum field
theory is also necessarily Lorentz violating~\cite{ref-greenberg}.
The SME was constructed precisely to provide a systematic approach, via an effective field theory framework
for understanding searches for violations of the fundamental Lorentz and CPT symmetries.

As the SME is an effective field
theory, it contains towers of operators of increasing mass dimension, with the contributions from
higher-dimensional operators assumed to be further suppressed under most circumstances.
The minimal SME is a restricted subtheory of the SME---that which is expected to be
renormalizable, because it satisfies the usual conditions that ensure renormalizability for
the standard model, by containing only the finite number of local, Hermitian, gauge-invariant
operators which can be built out of the known standard model fermion and boson field that are of
dimension four or less. The terms in the minimal SME Lagrange density are, in structure, very similar to
those appearing the usual standard
model; the key difference is that the additional Lorentz-violating operators each have one or more
uncontracted Lorentz indices. The coefficients multiplying those operators then become observable as 
components of preferred background vectors or tensors.
In many situations, the minimal SME is the most
natural test theory for analyzing the results of experimental searches for CPT or Lorentz violation.
In general, the SME may be thought up as a 
structure which enable scientists to translate the absence of experimentally observed signs of
Lorentz violation into constraints on other meaningful physical quantities to be measured~\cite{ref-schreck7}.

Here we shall be considering a Lorentz noninvariant theory along these lines. 
Many elementary aspects of the general SME remain somewhat mysterious and are open opportunities for
prospective research. Field theories with atypical properties can, even if those field theories are
never likely to be physically realized, provide useful theoretical laboratories for understanding how
quantum theories in general can behave; pure massless quantum electrodynamics in $1+1$ dimensions is unlikely
to exist physically, but that did not prevent the Schwinger model~\cite{ref-schwinger3,ref-callan1} from
educating us about the fact that there may sometimes
be an infinite number of vacua with different topologies.
Gauge invariance properties and radiative corrections to the SME effective field theory have, as a result,
achieved a plethora of theoretical
attentions~\cite{ref-coleman,ref-jackiw1,ref-chung1,ref-victoria2,ref-kost4,ref-altschul2-PV,ref-altschul37}.
The minimal SME, which includes Lorentz-violating field operators of mass dimension two, three, and four
will be the domain for our present analysis. Current physical experiments at relatively low energies are useful
for testing precisely this low-energy limit of the full SME. Since many low-energy experiments may be made
extremely precise, they may be more likely than data collected closer to the Planck scale to provide
empirical proof of a violation of the symmetries of the current standard model. 

We are setting out now to find extensions of our previous work~\cite{ref-alt+oco},
through examination of major aspects of Lorentz-violating quantum field theories with emphasis on the
behaviors of the modified particle momenta and energies~\cite{ref-altschul4}, and how they affect decay
kinematics in an unusual field theory.
We shall look specifically at three-body decay rates with Lorentz violation, doing computations with
a modified outgoing phase space.

Well-developed discussions of certain scattering cross sections, decay rates, and radiation emission rates
in the presence of Lorentz symmetry breaking exist in the literature, and kinematic considerations are often
of uppermost importance. Many of the modifications to the
standard theory that are necessary are described in Ref.~\cite{ref-kost5}; as an example, the authors calculate
the cross section for pair annihilation in the limit where particle momentum is greater than mass but still
far below the Planck energy. In Ref.~\cite{ref-altschul2}, the Compton scattering cross section is computed
in the presence of a Lorentz- and CPT-breaking background affecting the electron. In that case, since
the Lorentz violation breaks the spin degeneracy of the external state, the velocity and phase space factors
in the kinematics depend on the particles' spins; therefore Casimir’s trick for polarization sums does not
work for calculating the unpolarized cross section---which means the modified kinematics cause
a fundamental change in how the rate for the process has to be calculated. The cross section needs to be
determined using a basis of explicit polarization states for all the incoming and outgoing particles, which
can be quite complicated, even with a single isotropic but Lorentz-violating $b_{\mu}$ term added to the
action. The scattering cross section differs greatly from the Klein-Nishina formula, and despite the complexity,
the calculation can be performed nonperturbatively in $b_{\mu}$, making it possible to observe that the
cross section actually diverges at low photon energies---in sharp contrast to the predictions of various
low-energy theorems that assume Lorentz invariance.
In the electromagnetic sector of the minimal SME, there may exist a Chern-Simons-like Lorentz-violating
term. Besides being of intense theoretical interest because of its unusual gauge symmetry properties and
their relationship to radiative corrections, this term modifies in the dispersion relations for electromagnetic
waves (in a polarization-dependent fashion). This opens up the possibility of a Cerenkov-type effect in
vacuum, which is discussed in Ref.~\cite{ref-lehnert2}, which calculates the exact radiation rate using
the modified field equations and shows that they essentially agree with the phase space estimate. The
phase space estimate is calculated using a standard matrix element, combined with an evaluation of the available
outgoing phase space for the single almost-circular polarization mode into which radiation is kinematically
allowed. These kinematic considerations can be extended to a broad class of SME coefficients in the
charged fermion sector as well~\cite{ref-altschul12,ref-schreck4}.

With a similar methodology, we shall here be calculating decay rates in a simple
model of a three-body decay,
with a parent particle such as a kaon or $\eta$ meson decaying into three identical pions, in the presence
of Lorentz-violating but CPT-preserving $c_{\mu\nu}$ terms that modify the kinematics for the daughters.
As well as outlining the calculation of a total decay rate, we shall look at how the Lorentz violation
affects the boundaries of the allowed region of the Dalitz plot for the process.

The paper is organized as follows. In section~\ref{sec-theory}, we outline the specific type of SME theory
that we shall be using in our kinematics calculations. In section~\ref{sec-phase-space}, we consider
two specific textures for the Lorentz-violating background tensor $c_{\mu\nu}$, outlining the practicalities
of calculating a three-body decay rate with a modified space. Section~\ref{sec-dalitz} turns to the
determination of the boundaries of the Dalitz plot for the decay process, demarcating the kinematically
allowed region in the plane of the two independent mass-squared parameters. Finally, section~\ref{sec-concl}
summarizes our conclusion and discusses likely areas for future investigation.

\section{Lorentz-Violating Field Theory}
\label{sec-theory}

Phase space modifications due to Lorentz violation can increase the number of singularities in the
integrand and the boundary conditions in any integral for the decay rate $\Gamma$, and
the calculational approaches for specific decay processes  must be transformed in such ways as to
address these added complications. With anisotropic Lorentz-violating terms in the Lagrange density for
each outgoing particle, the angular integration may make calculating $\Gamma$ highly nontrivial,
since when we include such terms the particle energies depend on the angular orientations of
the outgoing particles' motions.
Therefore, before exploring more elaborate modified decay integrals in the
future, we shall consider in this paper a particularly simple model. The parent particle will decay into three
daughters with equal mass, vanishing spin, and identical Lorentz violation parameters.

Since we shall be focusing here on the
outgoing kinematics of a three-body decay, with the daughter particles all identical, we shall only need to
consider a sector of the minimal SME with a single field.
The Lagrange density, which involves the standard mass and kinetic terms from the quantum field theory,
as well as our Lorentz-violating background tensor, is 
\begin{equation}
{\cal L}=\frac{1}{2}(\partial^{\mu}
\pi)(\partial_{\mu}\pi)+c_{\mu\nu}(\partial^{\mu}\pi)
(\partial^{\nu}\pi)-\frac{1}{2}m^{2}\pi^{2}.
\label{eq-A}
\end{equation}
Here, $\pi$ is the spinless field of which the three daughter particles are excitations. The coefficients
$c_{\mu\nu}$ form a symmetric two-index tensor background, which describes violations of rotation and boost
symmetries, but without any breaking of CPT.
For a single real Klein-Gordon field, this is the only type of Lorentz violation possible in the minimal
SME context
The dispersion relation for the $\pi$ field---from which all the kinematic information may be derived---has
the relatively simple Lorentz-violating form
\begin{equation}
\left(g^{\mu\nu}+2c^{\mu\nu}\right)p_{\mu}p_{\nu}-m^{2}=0.
\label{eq-disp}
\end{equation}
The notation $k_{\mu\nu}$ for $2c_{\mu\nu}$ is often used in the context of spinless fields, with
$c_{\mu\nu}$ restricted to the physically analogous coefficients for fermions. However, since our analysis
here will focus purely on kinematics, the results should be equally valid for outgoing scalars and spinors,
provided the Lorentz violation is of the spin-independent $c_{\mu\nu}$ type. We therefore opt to use the
notation with $c_{\mu\nu}$ even for bosonic fields.

To canonically quantize a theory with $c_{\mu\nu}$-type Lorentz violation---whether the scalar theory
in (\ref{eq-A}) or its fermionic analogue---it is generally useful to rescale the
field and the mass parameter $m$. In the Klein-Gordon theory, that means a rescaling
so that the Lagrange density does not contain nonstandard double time
derivatives $c_{00}\partial^{0}\partial^{0}$. This may always be accomplished, because we may always subtract
a term proportional to $g_{\mu\nu}$ from $c_{\mu\nu}$, transferring the double time derivative from
the SME term to the 
conventional Klein-Gordon kinetic term in $(\ref{eq-A})$. Following this transformation and a rescaling
of $\pi$ and $m$, the Lagrange density still has the general form (\ref{eq-A}), albeit with
different values for the diagonal components of the background tensor. This rescaling, because it
changes the normalization of the quantized field, will appear in the matrix elements for any processes
involving the $\pi$ field. However, as we shall not be concerned here with the details of matrix element
calculations---basing our results purely on the kinematical corrections derived from the modified
dispersion relation (\ref{eq-disp})---we will not need to delve into that issue in detail.

In this paper, she shall focus on two specific textures for the $c_{\mu\nu}$ background tensor. A more
complete future analysis, covering all forms of anisotropy that could arise from a scalar-sector $c_{\mu\nu}$,
would consider sets of coefficients having different symmetry types separately, to see numerically
how those of each individual type
would shift the shape of the Dalitz plot for a specific decay at a known energy. When possible,
bounds on SME coefficients are typically reported in a standardized celestial equatorial coordinate system.
Time-dependent coordinate transformations may then be used to relate time-stamped laboratory data to 
coefficients in the standardized coordinates. The coefficients in celestial equatorial coordinates
will determine the amplitudes for time-dependent oscillations
in the shapes of the plots, at the Earth's sidereal rotation frequency $\omega_{\oplus}$ and also at
$2\omega_{\oplus}$~\cite{ref-bluhm4}.

\section{Decay Phase Space}
\label{sec-phase-space}

\begin{figure}
\centering
\includegraphics[angle=0,width=7cm]{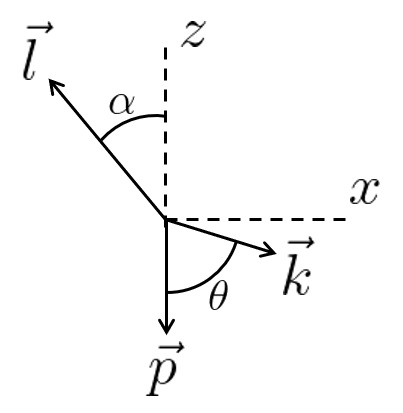}
\caption{Configuration of the $x$- and $z$-axes and the outgoing three-momenta in the plane of the decay.}
\label{fig-pkl}
\end{figure}

We begin with a general description of the differential decay rate for a channel with three identical spinless
particles in the final state; for definiteness, we shall take these daughter particles to be pions.
Beginning with a fairly general description, we shall make successive simplifying assumptions in order to
make calculations reasonably tractable. To start with, the differential decay rate for a channel in which
the three outgoing pion four-momenta are $p$, $k$, and $l$ is
\begin{equation}
\label{eq-dGamma}
d \Gamma=\frac{\left(2\pi\right)^{4}}{2}\frac{|\mathcal{M}|^{2}}{\sqrt{s}}\,d^{4}p\,d^{4}k\,d^{4}l\,
\delta(\xi_{p})\,\delta(\xi_{k})\,\delta(\xi_{l})\,\delta^{4}(P-p-k-l).
\end{equation}
This will form the integrand of the modified decay integral, with modified quadratic Lorentz-violating dispersion
relations visible in the first three Dirac $\delta$-functions. These enforce the energy-momentum relations
for each of the three pions in the final state. If the initial unstable particle is at rest, so that
$P^{\mu}=(\sqrt{s},\vec{0}\,)$, the three daughter three-momenta lie in a plane. We may therefore take
the spatial momentum of the first daughter pion to define the $-z$-direction,
$p^{\mu}=(E_{p},0,0,-|\vec{p}\,|)$,
with the other two daughter trajectories also lying in the $xz$-plane.
In the rest frame of the parent particle, $\theta$ is the angle between $\vec{p}$ and $\vec{k}$, and
$\alpha$ is the angle between $\vec{l}$ and $\hat{z}$. This configuration is shown in figure~\ref{fig-pkl}.
(Note that because of the anisotropy in $c_{\mu\nu}$, the three-momentum of a particle may point in
a slightly different direction from its velocity.)

Furthermore, we shall eventually assume that
$\sqrt{s}\gg m$, so that the outgoing mesons are ultrarelativistic. For the fermionic theory with
$c_{\mu\nu}$, which has a
slightly more intricate structure than the scalar theory, the ultrarelativistic limit results in some
additional simplifications. For the scalar theory, or for the ultrarelativistic Dirac theory,
the arguments of the
first three $\delta$-functions in the expression for $d\Gamma$ are
\begin{eqnarray}
\label{eq-xip}
\xi_{p} & = & \left(1+c_{00}\right)E_{p}^{2}-c_{03}\left|\vec{p}\,\right|E_{p}-
\left(1+c_{33}\right)\left|\vec{p}\,\right|^{2}-m^{2} \\
\label{eq-xik}
\xi_{k} & = & (1+c_{00})E_{k}^{2}+(c_{01}\sin\theta-c_{03}\cos\theta)|\vec{k}|E_{k} \nonumber\\
& & -(1-c_{11}\sin^{2}\theta-c_{33}\cos^{2}\theta+c_{13}\sin\theta\cos\theta)|\vec{k}|^{2}-m^{2} \\
\label{eq-xil}
\xi_{l} & = & (1+c_{00})E_{l}^{2}+(c_{01}\sin\alpha+c_{03}\cos\alpha)|\vec{l}\,|E_{l} \nonumber\\
& & -(1-c_{11}\sin^{2}\alpha-c_{33}\cos^{2}\alpha+c_{13}\sin\alpha\cos\alpha)|\vec{l}\,|^{2}-m^{2}.
\end{eqnarray}
The angular structure that arises from integration over all allowed momenta is extremely complicated if
the texture of $c_{\mu\nu}$ is allowed to contain all nine physically distinguishable parameters, describing
generic forms of rotation and boost invariance violations in the pion kinetic energy. So a simplifying
assumption is desirable, as our principal goal is to describe the complications that arise in evaluations
of the three-body decay phase space factor in the context of the SME. In the literature, as a way of reducing
a general symmetry violation problem to one with just a single parameter describing the extent of the Lorentz
violation, it is common to consider a theory that is isotropic (in the lab frame or some other natural frame,
such as the rest from of the cosmic microwave background; in that particular frame, only the Lorentz boost
symmetry is broken).

However, in a case like this one, the isotropic theory is actually too simple. With identical
ultrarelativistic
daughter particles, the Lorentz-violating modifications in an isotropic theory can be reduced to
a simple rescaling
of the conventional result. Actually, this is a fairly natural conclusion. The complexity of the $d\Gamma$ in
(\ref{eq-dGamma}) comes from the angular structure, and if there is no anisotropy $c_{0j}=0$ and $c_{jk}$ is
proportional to $\delta_{jk}$, which wipes out all the angular structure in
(\ref{eq-xip}--\ref{eq-xil}).
If there are no angular dependences, it does not meaningfully matter which
directions the daughter pions are moving; the only change is an overall rescaling of the dependence of
the particle energy on the spatial momentum. So we shall instead introduce a different simplifying
parameterization of the background tensor $c_{\mu\nu}$.

\subsection{Single-Coefficient Model}
\label{sec-single}

Analogous to an isotropic limit, we could set all the $c$ coefficients equivalent,
$c_{\mu\nu}=c\mathbf{M}_{4\times4}$,
where all the components of the matrix $\mathbf{M}_{4\times4}$ are equal to 1. A theory with a $c_{\mu\nu}$
background tensor of this form is certainly not Lorentz invariant; only if $c_{\mu\nu}\propto g_{\mu\nu}$
is the Lorentz symmetry unbroken. However, it can be slightly trickier to identify whether or not a
SME-type theory is isotropic in some particular preferred frame. In this case,
$c_{\mu\nu}=c\mathbf{M}_{4\times4}$ is not an isotropic
texture, as there is a clearly preferred direction, $(\hat{x}+\hat{y}+\hat{z})/\sqrt{3}$. So nontrivial
anisotropy is preserved, while still making the modifications to the conventional theory depend on just a
single dimensionless parameter $c$. The differential decay rate with this type of $c_{\mu\nu}$ is
\begin{eqnarray}
d\Gamma & = & \frac{\left(2\pi\right)^{4}}{2}\frac{|\mathcal{M}|^{2}}{\sqrt{s}}\,d^{4}p\,d^{4}k\,d^{4}l\,
\delta[(1+c)E_{p}^{2}-c|\vec{p}\,|E_{p}-(1+c)\left|\vec{p}\,\right|^{2}-m^{2}] \nonumber\\
& & \times\,\delta\left[(1+c)E_{k}^{2}+c(\sin\theta-\cos\theta)|\vec{k}|E_{k}
-(1-c+c\sin\theta\cos\theta)|\vec{k}|^{2}-m^{2}\right] \nonumber\\
& & \times\,\delta\left[(1+c)E_{l}^{2}+c(\sin\alpha+\cos\alpha)|\vec{l}\,|E_{l}
-(1-c+c\sin\alpha\cos\alpha)|\vec{l}\,|^{2}-m^{2}\right] \nonumber\\
& & \times\,\delta^{4}(P-p-k-l).
\end{eqnarray}
However, it is possible to simplify this expression even somewhat further while still
retaining nontrivial angular anisotropy. The calculation steps and final expression are simplified greatly
if we set $c_{13}=c_{31}=0$. A great deal more in-depth analysis of the physical process will be
needed if the mixing
of the $x$- and $z$-components is to be kept in our calculation. The angular analysis should be made quite
complex if this extra dependence is to be considered, so the decay with this term nonzero will be evaluated
later. For now we drop this term because without it a simple cancellation arises without angular
approximations. This is, we set $c_{\mu\nu}=c\mathbf{M}'_{4\times4}$, where $\mathbf{M}'_{4\times4}$
agrees with $\mathbf{M}_{4\times4}$, except that
$\left(\mathbf{M}'_{4\times4}\right)_{13}=\left(\mathbf{M}'_{4\times4}\right)_{31}=0$. It is common practice, when
determining the maximum sensitivity of searches for Lorentz violation, to consider only restricted subsets of
SME coefficients, rather than all of them at once. Operators with different parities under discrete
symmetries will not mix under quantum corrections or through other modes of mediation. Since $c_{13}$ has
different behavior under time reversal, charge conjugation, and spatial reflections along the $x$- and $z$-axes
from any other minimal SME operator, it may be safely omitted. If $c_{13}=0$ in the fundamental action,
it will be protected from being generated by any indirect effects.

With this modification,
the differential decay rate for the channel is
\begin{eqnarray}
d\Gamma & = & \frac{\left(2\pi\right)^{4}}{2}\frac{|\mathcal{M}|^{2}}{\sqrt{s}}\,d^{4}p\,d^{4}k\,d^{4}l\,
\delta[(1+c)E_{p}^{2}-c|\vec{p}\,|E_{p}-(1+c)\left|\vec{p}\,\right|^{2}-m^{2}] \nonumber\\
& & \times\,\delta\left[(1+c)E_{k}^{2}+c(\sin\theta-\cos\theta)|\vec{k}|E_{k}
-(1-c)|\vec{k}|^{2}-m^{2}\right] \\
& & \times\,\delta\left[(1+c)E_{l}^{2}+c(\sin\alpha+\cos\alpha)|\vec{l}\,|E_{l}
-(1-c)|\vec{l}\,|^{2}-m^{2}\right]\,\delta^{4}(P-p-k-l). \nonumber
\end{eqnarray}

Continuing with the purely kinematical calculation, we may integrate over the momenta of the third particle.
The $\delta$-function that enforces overall momentum conservation sets
$\vec{l}=-(\vec{p}+\vec{k})$, leaving 
\begin{eqnarray}
d\Gamma & = & \frac{\left(2\pi\right)}{2}\frac{|\mathcal{M}|^{2}}{\sqrt{s}}
\,d^{4}p\,d^{4}k\, dE_{l}\sum_{\pm}\frac{\delta(E_{l}-\omega_{l}^{\pm})}{2|\omega_{l}^{\pm}|}
\delta\!\left[(1+c)E_{p}^{2}-c|\vec{p}\,|E_{p}-(1+c)|\vec{p}\,|^{2}-m^{2}\right] \nonumber\\
& & \times\,\delta\left[(1+c)E_{k}^{2}+c(\sin\theta-\cos\theta)|\vec{k}|E_{k}
-(1-c)|\vec{k}|^{2}-m^{2}\right] \nonumber\\
& & \times\,\delta\left[\sqrt{s}-E_{p}-E_{k}-E_{l}(\vec{l}=-\vec{p}-\vec{k}\,)\right].
\end{eqnarray}
Here, $\sum_{\pm}\delta(E_{l}-\omega_{3}^{\pm})$ appears because of the mathematical possibility of
$E_{l}$, which is determined by a quadratic energy-momentum relation, could potentially takes roots
$\omega_{l}^{\pm}(\vec{l}\,)$ of either positive or negative sign. Physically, however,
obviously only the positive-energy root
\begin{equation}
2\omega_{l=-p-k}^{+}=\sqrt{4(1+c)\left[(1-c)|\vec{p}+\vec{k}|^{2}+m^{2}\right]
+c^{2}\left[|\vec{p}\,|
-|\vec{k}|(\sin\theta-\cos\theta)\right]^{2}}
\end{equation}
contributes to the decay rate.

Conversion of the energy-preserving $\delta$-functions for the other two particles
and dropping all but the positive, physical branches of the dispersion relations
gives two more energy square root terms in the denominator
\begin{eqnarray}
\label{eq-dGamma-c2}
d\Gamma & = & \frac{1}{2(2\pi)^{5}}\frac{|\mathcal{M}|^{2}}{\sqrt{s}}\frac{d^{3}p\,d^{3}k}
{\sqrt{4(1+c)\left[(1-c)|\vec{p}+\vec{k}|^{2}+m^{2}\right]
+c^{2}\left[|\vec{p}\,|
-|\vec{k}|(\sin\theta-\cos\theta)\right]^{2}}} \nonumber\\
& \times & \frac{\delta\left[\sqrt{s}-E_{p}-E_{k}-E_{l}\left(\vec{l}=-\vec{p}-\vec{k}\right)\right]}
{\sqrt{|\vec{p}\,|^{2}(4-3c^{2})+4(1+c)m^{2}}
\sqrt{|\vec{k}|^{2}(4-3c^{2}-2c^{2}\cos\theta\sin\theta)+4(1+c)m^{2}}}.
\end{eqnarray}
Since one of our principal focuses will be understanding the nature of any anisotropic effects due to the
Lorentz violation, we could make a further approximation at this point. Without the inclusion of
$\mathcal{O}(c^{2})$ terms in $d\Gamma$, there is no nonstandard angular dependence. However, since terms
at this order are doubly small, it is of less profit to maintain them in our expressions when the are
not specifically anisotropic. We could therefore drop terms of $\mathcal{O}(c^{2})$ that do not also
depend on $\theta$; this entails, for example, the approximation $4-3c^{2}\approx 4$. With this done,
the approximate energies that appear in (\ref{eq-dGamma-c2}) would become
\begin{eqnarray}
(1+c)E_{p} & \approx & \sqrt{|\vec{p}\,|^{2}+(1+c)m^{2}} \\
(1+c)E_{k} & \approx & \sqrt{|\vec{k}|^{2}+(1+c)m^{2}-\frac{c^{2}}{2}|\vec{k}|^{2}\sin\theta\cos\theta} \\
(1+c)E_{l} & \approx & \sqrt{|\vec{p}+\vec{k}|^{2}+(1+c)m^{2}-\frac{c^{2}}{2}\left[|\vec{p}\,||\vec{k}|
(\sin\theta-\cos\theta)-|\vec{k}|^{2}\sin\theta\cos\theta\right]}.
\end{eqnarray}
However, we shall, for now, continue without this explicit simplification.

The next major step in the calculation is to deal with the energy conservation $\delta$-function,
which depends on the angles.
This remaining $\delta$-function in $d\Gamma$ has the complicated argument
$\Delta=\sqrt{s}-E_{p}-E_{k}-E_{l}$, with the overall form,
\begin{eqnarray}
\delta(\Delta) & = & \delta\left(\sqrt{s}-\frac{1}{2(1+c)}
\left\{\sqrt{|\vec{p}\,|^{2}(4-3c^{2})+4(1+c)m^{2}}\right.\right. \nonumber\\
& & \left.\left.+\sqrt{|\vec{k}|^{2}(4-3c^{2}-2c^{2}\cos\theta\sin\theta)+4(1+c)m^{2}}\right.\right. 
\nonumber\\
& & \left.\left.+\sqrt{4(1-c^{2})|\vec{p}+\vec{k}|^{2}+4(1+c)m^{2}+
c^{2}[|\vec{p}\,|-|\vec{k}|(\sin\theta-\cos\theta)]^{2}}\right\}\right).
\end{eqnarray}
Implementation of this $\delta$-function will entail multiplication by
$|\partial\Delta/\partial\theta|^{-1}$.
Explicitly, this angular derivative is
\begin{equation}
\left|\frac{\partial\Delta}{\partial\theta}\right|=
\frac{1}{2(1+c)}\left\{
\frac{4|\vec{p}\,||\vec{k}|\sin\theta}{E_{l}}
+c^{2}\left[\frac{|\vec{k}|^{2}\cos2\theta+|\vec{p}\,||\vec{k}|
(\cos\theta-3\sin\theta)}{E_{l}}
+\frac{|\vec{k}|^{2}\cos2\theta}{E_{k}}\right]\right\}.
\end{equation}
Note that since the Lorentz-invariant term is always positive, the signs of the small Lorentz-violating
corrections are unimportant to the determination of the absolute value.

After dividing the integrand in $d\Gamma$
by $|\partial\Delta/\partial\theta|$, the resulting integral is manageable
at first order in $c$. The terms which depend on $\theta$ cancel after the term is divided into the integral
and then the whole integrand approximated to first order in $c$.  This avoids the complicated integration
that would be necessary if the angular roots needed to be plugged back into $\theta$-dependent terms
remaining in the integrand.

At this point, it is not possible to go further using only facts about the outgoing particle kinematics.
For a generic process, the matrix element $i\mathcal{M}$ will also be a function of $\vec{p}$ and
$\vec{k}$ (as well as the constrained momentum $\vec{l}$). The dependence can be both on
the magnitudes of the three outgoing pion momenta and the angles between them. For the purpose of continuing
our kinematical analysis, we must therefore make choice of what matrix element to use. To simplify things
to the greatest extent possible---so as to keep the focus on what effects may be attributed purely to
daughter particle kinematics---we shall opt for the simplest possible choice: a constant $|\mathcal{M}|^{2}$,
such as might arise from a four-meson contact interaction of the form
$\mathcal{L}_{I}\propto\lambda(\eta\pi^{3})$.
Henceforth, the $|\mathcal{M}|^{2}$ appearing in our expressions will be treated as a momentum-independent
constant. This constant may generically include additional factors of $c$; in particular, the rescaling
of the $\pi$ field to remove $c_{00}\partial^{0}\partial^{0}$ terms mentioned above can introduce
$\mathcal{O}(c)$ momentum-independent corrections to the matrix element.

Although the approximation of constant $|\mathcal{M}|^{2}$ may seem rather drastic, it is not necessarily
that unrealistic. The leading-order strong interaction
matrix element for $\eta\rightarrow3\pi^{0}$ in chiral perturbation
theory is actually a momentum-independent constant~\cite{ref-ditsche} (although because the process
violates isospin, the electromagnetic contribution to the matrix element is of similar size).

With the angular dependence of the matrix element squared known (and in our toy model case, trivial), it
is possible to perform the angular integrations explicitly, at least to leading order in the Lorentz violation.
The integration over the angular variables that the integrand does not depend upon (meaning over the
direction of $\vec{p}$, and the azimuthal angle of $\vec{k}$)
produces a factor a factor of $2(2\pi)^{2}$. The last angular integration over $\theta$
gives new factors that yield
\begin{equation}
d\Gamma = \frac{1}{64\pi^{2}}\frac{|\mathcal{M}|^2}{\sqrt{s}}\,d|\vec{p}\,|d|\vec{k}|\,
\frac{|\vec{p}\,||\vec{k}|(m^{4}+m^{2}\left[|\vec{p}\,|^2(1+\frac{c}{2})+|\vec{k}|^2(1+\frac{c}{2})
\right]+|\vec{p}\,|^2|\vec{k}|^2(1+c)}{(|\vec{k}|^{2}+m^{2})^{3/2}(|\vec{p}\,|^{2}+m^{2})^{3/2}}.
\label{eq-analog}
\end{equation}
To simplify the remaining integration further, we can convert the integral in momentum to an integral over
the unmodified on-shell energies  
\begin{eqnarray}
p^{(0)} & = & \sqrt{|\vec{p}\,|^{2}+m^{2}} \\
k^{(0)} & = & \sqrt{|\vec{k}|^{2}+m^{2}}.
\end{eqnarray}
The integral in terms of these conventional energies simplifies to
\begin{equation}
d \Gamma = \frac{1}{64\pi^2}\frac{|\mathcal{M}|^2}{\sqrt{s}}\,dp^{(0)}\,dk^{(0)}\,
\left\{1+c-\frac{c}{2}\left[\frac{m^{2}}{(p^{(0)})^{2}}+\frac{m^{2}}{(k^{(0)})^{2}}\right]\right\}.
\end{equation}
Although we have not yet integrated over the magnitudes of the
spatial momenta [or, equivalently, over $p^{(0)}$
and $k^{(0)}$], but only the associated angles, any momentum-dependent matrix element $i\mathcal{M}$
would need to be symmetric in $\vec{p}$, $\vec{k}$, and $\vec{l}=-\vec{p}-\vec{k}$; and the symmetric
dependence on $\vec{l}$ means that if $i\mathcal{M}$ depends on the magnitudes of the three-momenta,
it will also unavoidably depend on $\theta$. So only with a completely constant matrix element may we
reach this and later expressions.

We shall now impose the ultrarelativistic limit mentioned earlier. This allows us to neglect the mass
scale $m$, except when it is needed to regulate logarithmic divergences in the infrared. With the
$m\approx0$ approximation, the first momentum integration is
\begin{equation}
d\Gamma = \frac{1}{64\pi^2}\frac{|\mathcal{M}|^2}{\sqrt{s}}\,dp^{(0)}
\int\limits_{\frac{\sqrt{s}(1+c)}{2}-p^{(0)}}^{\frac{\sqrt{s}(1+c)}{2}}dk^{(0)}
\left\{1+c-\frac{c}{2}\left[\frac{m^{2}}{(p^{(0)})^{2}}+\frac{m^{2}}{(k^{(0)})^{2}}\right]\right\}.
\end{equation}
We cannot use 0 as the lower bound in the subsequent $p^{0}$ integration, because of an infrared
$\log(0)$ divergence. Instead, the lower bound must be cut off by the physical
minimal value of $p^{(0)}$, which is $m$. This leaves, after integration
over the $k^{(0)}$ energy,
\begin{equation}
\Gamma = \frac{1}{64\pi^2}\frac{|\mathcal{M}|^2}{\sqrt{s}}
\int\limits_{m}^{\frac{\sqrt{s}(1+c)}{2}}dp^{(0)}
\left[p^{(0)}+c\left(p^{(0)}-\frac{m^2}{2p^{(0)}}-\frac{2p^{(0)}m^{2}}{s-2\sqrt{s}p^{(0)}}\right)\right].
\end{equation}
The calculation results in the ultimate decay rate 
\begin{equation}
\label{eq-Gamma-c13=0}
\Gamma=\frac{|\mathcal{M}|^2}{512\pi^2\sqrt{s}}\left[s(1+3c)-4m^{2}
+4c m^{2}\log\left(\frac{2cm}{\sqrt{s}-2m}\right)\right].
\end{equation}

The form of the $c$ dependence in (\ref{eq-Gamma-c13=0}) is instructive. Note that although we expanded
various expressions to $\mathcal{O}(c)$, the final expression's dependence on $c$ is slightly different,
containing a term of size
$\mathcal{O}(c\log c)$. Naively, this looks like a slightly stronger dependence on the coefficient
$c$. However, the $\mathcal{O}(c\log c)$ term is also proportional to another small quantity, $m^{2}/s$.

\subsection{Model with Nonzero $c_{13}$}

If instead we keep all terms in $c\mathbf{M}_{4\times4}$, including the $c_{13}$ term,
things grow substantially more complicated, and angular approximations will be needed to get
tractable analytical expressions.
To ease the number of terms in the integrand for the decay rate, we will eventually use
a series approximation that can be performed in $\theta$ to compute the angular part of the integral.

With the added angular dependence in the square root terms from the modified dispersion relations,
two of the three pions dispersion relation become modified to include first-order sinusoidal terms.
The analog of (\ref{eq-analog}), before $\theta$-integration but with keeping the additional anisotropy
term, is
\begin{eqnarray}
d\Gamma & = & \frac{1}{128\pi^{3}}\frac{|\mathcal{M}|^2}{\sqrt{s}}\,d|\vec{p}\,|d|\vec{k}|\,d\theta\,
\frac{|\vec{k}|\csc\theta}{|\vec{p}\,| (|\vec{k}|^2+m^2)^{3/2}(|\vec{p}\,|^2+m^2)^{3/2}} \nonumber\\
& & \times\bigg(
4|\vec{p}\,|^{2}\left\{(2+c)|\vec{p}\,|^{2}m^{2}+2m^{4}+
|\vec{k}|^{2}\left[2(1+c)|\vec{p}\,|^{2}+(2+c)m^{2}\right]\sin\theta\right\} \nonumber\\
& & +\,c(5|\vec{k}|^2+4m^{2})(|\vec{p}\,|^{2}+m^{2})|\vec{p}\,|^{2}\cos\theta
+4c|\vec{k}||\vec{p}\,|(|\vec{p\,}|^{2}+m^{2}) \nonumber\\
& & \times
\left[|\vec{k}|^{2}+m^{2}-\sqrt{|\vec{k}|^{2}+m^{2}}\sqrt{|\vec{k}|^{2}+|\vec{p}\,|^{2}+m^{2}+2|\vec{p}\,|
|\vec{k}|\cos\theta}\right]\cos 2\theta \nonumber\\
& & +\, c|\vec{p}\,|^{2}|\vec{k}|^{2}\left[|\vec{p}\,|^{2}+m^{2}\right]\cos3\theta\bigg).
\label{eq-Gamma-all-c-angles}
\end{eqnarray}
Because of the $\csc\theta$, the $c$-dependent part of the
the above integral clearly diverges when the three daughter pions particles are colinear.
This appears to be indicative of the previously noted fact that there may be a nonperturbative
dependence on $c$ where logarithms arise in the decay rate $\Gamma$; the $\mathcal{O}(c)$
approximations that we have made are inadequate near the poles at $\theta=0$ and $\pi$, where
$\cos n\theta=(-1)^{n}$.

We could now regulate the integral equation by analytical continuation
and use regulation techniques like asymptotic approximation~\cite{ref-zeta}, to account for the
unusual Lorentz-violating behavior of the $\theta$ integration boundaries.
However, if we merely continue the calculation by taking the indefinite integral over $\theta$ and using the
fact that the $\theta$ integration region is symmetric around $\theta=\frac{\pi}{2}$, we note that the
functions $\cos\theta$ and $\cos3\theta$ are odd in $\vartheta=\theta-\frac{\pi}{2}$. Therefore, with
the $\theta$-independent matrix element squared $|\mathcal{M}|^{2}$, the terms
with $\cos\theta$ and $\cos3\theta$ will not make any $\mathcal{O}(c)$ contributions to the fully integrated
$\Gamma$.

If the matrix element happened to be strongly peaked around $\vartheta=0$, we might then further
approximate, to first order in $\vartheta$, that $\cos 2\theta=-1$. However, this is not a good
approximation in the identical particle model we are using here.
On the other hand, without the
approximation, the angular integration becomes unavoidably tangled up with the integrals
over $|\vec{p}\,|$ and $|\vec{k}|$. Going further and including the $c$-dependences of the limits of these
latter integrations produces expressions too complicated to provide analytical insights, except insofar as
we can see that it shares a somewhat similar logarithm structure to what we previously found in
(\ref{eq-Gamma-c13=0}). If the expansion is made around $\theta=\frac{\pi}{4}$,
an interesting expression for the decay rate arises,
\begin{eqnarray}
\Gamma &=& \frac{|\mathcal{M}|^2}{512\pi^2\sqrt{s}}\left\{\sqrt{\frac{s-4m^{2}}{s-3m^{2}}}(4s -12m^{2})
+4m^{2} \log\left({\frac{\sqrt{s-3m^{2}}-\sqrt{s-4m^{2}}}{m}}\right)\right. \nonumber\\
& & +\left.c\left[\sqrt{\frac{s-4m^{2}}{s-3m^{2}}} (7m^{2} -3s)
+4m^{2} \log\left({\frac{\sqrt{s-3m^{2}}-\sqrt{s-4m^{2}}}{m}}
\right)\right]\right\}.
\label{eq-Gamma-c13-nonzero}
\end{eqnarray}

To complete the integration without any angular approximations, we need to find the poles
in the boundary equation; we may do this by using (\ref{eq-boundary}) below. 
After insertion of the angle into (\ref{eq-Gamma-all-c-angles}), even when approximating to
$\mathcal{O}(c)$ the integral is extremely long. 
If the approximation is further made of working only to first or second order in the mass $m$,
the integral over the daughter momenta becomes more tractable; however,
combined with the $\mathcal{O}(c)$ approximation (even with using the modified upper and lower bounds
for the integration) the first-order dependence on $m$ vanishes. 
Thus, to obtain a non-vanishing modification to the decay rate,
the integral over on-shell daughter particle  energies must be computed to at least
$\mathcal{O}(m^{2})$.
The integration over one daughter energy yields a hugely complicated expression but with no
explicit Lorentz violation
terms remaining at leading order. However, after integration over the second on-shell energy, there is
a $c$ dependence, and the full decay rate at this order is 
\begin{equation}
\label{eq-Gamma-c13 not 0}
\Gamma=\frac{|\mathcal{M}|^2}{512\pi^2\sqrt{s}}\left[s(1+2c)
+4cm^{2}+4cm^{2}\log\left(\frac{2mc}{\sqrt{s}-2m}\right)\right].
\end{equation}

\section{Modified Dalitz Plots}
\label{sec-dalitz}

For a three-body decay, studied in the center of mass frame, it makes sense to express the kinematics using
a Dalitz plot.
When all three outgoing particles are identical, with or without Lorentz violation, the Dalitz plot is
symmetric with respect to the two coordinate axes, since forming the scalar quantities $m_{13}^{2}$ and
$m_{23}^{2}$ drops the leading order directional information. However, with non-identical daughters, the
convex outline may take a more asymmetrical shape. At higher energies,
as all particles become strongly relativistic, the shape approaches triangularity.

The interior of the allowed region of the plot is populated with a histogram of decay rates
for different values of the parameters. The difference between a flat, uniform histogram and
what is actually observed reveals the character of the dynamical matrix element. Dalitz plots
are especially useful for displaying or recognizing information about final state interactions.
If the invariant mass squared for two outgoing particles corresponds to a possible resonance intermediate,
there will be a line of resonance enhancement crossing the plot, something that can be easy to recognize
in a plot of experimental data.

So we shall now explicitly calculate the modified boundary of the kinematically allowed region
that arises from the inclusion of the Lorentz violation coefficients.
This boundary condition arises from integration over the energy conservation $\delta$-function;
with a first order approximation in $c$ and using
the $c_{13}=0$ model from section~\ref{sec-single}, the boundary condition is
\begin{eqnarray}
\sqrt{s}(1+c)& - & \left(\sqrt{|\vec{p}\,|^{2}+m^{2}}+\sqrt{|\vec{k}|^{2}+m^{2}}\right)
-\frac{c}{2}\left(\frac{m^{2}}{\sqrt{|\vec{k}|^{2}+m^{2}}}+\frac{m^{2}+2|\vec{p}\,|^{2}}
{\sqrt{|\vec{p}\,|^{2}+m^{2}}}\right) \nonumber\\
& \lessgtr & \sqrt{|\vec{k}|^{2}+|\vec{p}\,|^{2}+m^{2}\mp2|\vec{p}\,||\vec{k}|}
+\frac{cm^{2}}{2\sqrt{|\vec{k}|^{2}+|\vec{p}\,|^{2}+m^{2}\mp2|\vec{p}\,||\vec{k}|}}.
\label{eq-boundary-ineq}
\end{eqnarray}
If $c_{13}$ is not neglected, the expression (\ref{eq-boundary-ineq}) will remain unchanged;
the presence of $c_{13}$ only brings in
terms which involve $\sin\theta$, and on the kinematic boundaries these terms vanish.
Since the Lorentz violation coefficients $c$ are small and three-pion decays typically have relativistic
outgoing particles, $m^{2}\ll |\vec{p}\,|^{2},\,|\vec{k}|^{2},\,\sqrt{s}$,
we could simplify this expression by dropping
those terms that are small in both $c$ and $m^{2}$. With this logic, neglecting any term which multiplies both
$c$ and $m^{2}$, the boundary condition is simplified to
\begin{equation}
\sqrt{s}(1+c)-\sqrt{|\vec{p}\,|^{2}+m^{2}}-\sqrt{|\vec{k}|^{2}+m^{2}}-c|\vec{p}\,|
\lessgtr \sqrt{|\vec{k}|^{2}+|\vec{p}\,|^{2}+m^{2}\mp2|\vec{p}\,||\vec{k}|}
\label{eq-boundary}
\end{equation}
which is almost as simple as the conventional boundary condition with vanishing $c$.

The Dalitz plot is affected by the Lorentz violation through the modified boundary condition.
The boundary line for this plot can be plotted in terms of the unmodified invariant mass-squared
parameters, which are given in terms of the unmodified single-particle energies,   
\begin{eqnarray}
p^{(0)} & = & \sqrt{|\vec{p}\,|^{2}+m^{2}}=\frac{s+m^{2}-m_{13}^2}{2\sqrt{s}} \\
k^{(0)} & = & \sqrt{|\vec{k}|^{2}+m^{2}}=\frac{s+m^{2}-m_{23}^2}{2\sqrt{s}},
\end{eqnarray}
which are plugged into the boundary (\ref{eq-boundary}).

\begin{figure}[h]
\centering
\includegraphics[angle=0,width=8.5cm]{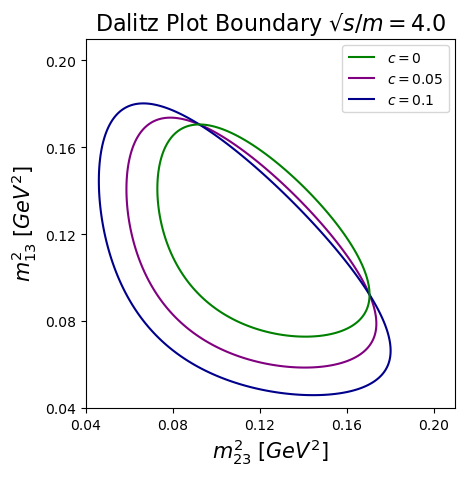}
\caption{Shape of the Dalitz plot for the $\sqrt{s}/m$ value corresponding to
$\eta\rightarrow3\pi^{0}$. The standard outline is shown, along with outlines for two nonzero values
of the Lorentz violation parameter $c$.}
\label{fig-Dalitz1}
\end{figure}

This Dalitz plot boundary line is given by a modified quartic equation in the invariant mass squared
parameters $m_{13}^{2}$ and $m_{23}^{2}$. With a $c=0$ and all the daughter particle masses the same the
boundary equation is 
\begin{equation}
0=(m_{13}^{2})^{2} m_{23}^{2}+ m_{13}^{2}(m_{23}^{2})^{2} -
(3 m^{2} + s)m_{13}^{2}m_{23}^{2}+ (m^{2} - s)^{2} m^{2}.
\label{eq-regb}
\end{equation}
With the inclusion of Lorentz violation term, the exact quartic is extremely intricate,
but to leading order in $c$ it takes the tractable form
\begin{eqnarray}
0 & = & (m_{13}^{2})^{2} m_{23}^{2}+ m_{13}^{2}(m_{23}^{2})^{2} -
(s+3 m^{2})m_{13}^{2}m_{23}^{2}+ (s-m^{2})^{2} m^{2} \nonumber\\
& & +\,\frac{c}{2s} \{(s-m^2 + m_{13}^2 + m_{23}^2)(2m^2 - m_{13}^2 - m_{23}^2)\nonumber\\
& &  [s^{2}-m^{4} - m_{23}^2 m_{13}^2 -(s- m^{2})(m_{13}^2 + m_{23}^2)]\}
\label{eq-modb}
\end{eqnarray}

The boundary line is plotted in the $m_{23}^{2}$-$m_{13}^{2}$ plane. Setting the quantity $c$ 
to zero as in (\ref{eq-regb}) naturally gives the standard outline of the plot. In this case, the
elongation of the curve depends only on the available decay energy compared with the common
mass of the daughters, $\sqrt{s}/m$. When $c$ is increased to a nonzero value the boundary plot
shifts according to (\ref{eq-modb}).

\begin{figure}[h]
\centering
\includegraphics[angle=0,width=8.5cm]{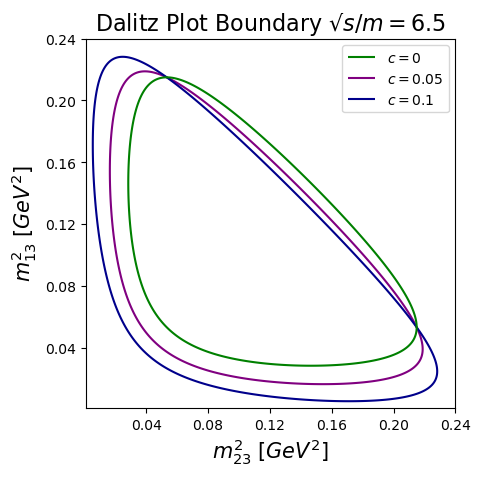}
\caption{Shape of the Dalitz plot for a $\sqrt{s}/m=6.5$ value.}
\label{fig-Dalitz2}
\end{figure}

This is linked to the fact that under Lorentz transformation, the allowed physical region will be
rotated and shifted relative to a frame with no Lorentz violation.
Increasing the $c$ term in the modified plot is associated with enlarging the physically allowed
parameter region,
as well as generally shifting it towards smaller values of $m_{13}^{2}$, $m_{23}^{2}$, and
(redundantly) $m_{23}^{2}$.

\begin{figure}
\centering
\includegraphics[angle=0,width=8.5cm]{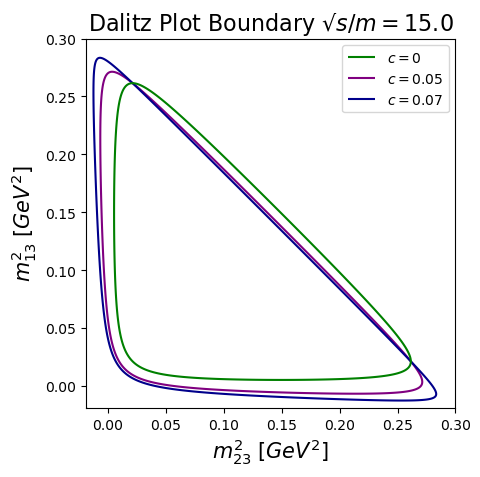}
\caption{Shape of the Dalitz plot for an ultrarelativistic value of $\sqrt{s}/m=15.0$.}
\label{fig-Dalitz3}
\end{figure}

Figure~\ref{fig-Dalitz1} shows the effect of $c$ on the shape of the Dalitz plot for a decay
with $\sqrt{s}/m=4.0$, very close to the value of 3.97 for the physical decay
$\eta\rightarrow3\pi^{0}$.
Figures~\ref{fig-Dalitz1}--~\ref{fig-Dalitz3} show the Dalitz plot boundaries for values of the
$c_{\mu\nu}$ coefficients at the $10^{-2}$--$10^{-1}$ level. These relatively large values are chosen
in part to enhance the visual contrast. However, although there are suggestive indications that they might
actually be made tighter~\cite{ref-kamand1}, direct two-sided bounds on the anisotropic quark-sector coefficients
that underlie the $c$ coefficient involved
here---coming from deep inelastic scattering~\cite{ref-abt} and Drell-Yan~\cite{ref-lunghi1}
process data---are only at $10^{-4}$--$10^{-1}$ levels~\cite{ref-tables}. So the study of these
kinds of Dalitz plots has the potential to be directly useful even at the levels shown in the figures.

Figure~\ref{fig-Dalitz1} shows that increasing the strength of the Lorentz violation (in this specific model),
shifts the kinematically allowed region towards smaller values of $m_{23}^2$ and $m_{13}^2$---in a symmetric
fashion, because the daughter particles are identical. Precision measurements of this particular decay
are already recognized to be important, since the matrix element, calculated using chiral perturbation
theory, depends directly on the isospin-violating
up-down mass difference, $m_{d}-m_{u}$~\cite{ref-anisovich,ref-bijnens,ref-kampf1}. Measurements
of this and similar processes continue to be used for state-of-the-art measurements of this important
parameter in low-energy flavor physics~\cite{ref-colangelo2,ref-ablikim}.

With time-stamped and pion tracking data, it would also be possible
to search these data sets for evidence of Lorentz violation. This would entail constructing separate
Dalitz plots for different bins of sidereal time, so see whether the kinematically allowed regions
shifts around in the $m_{23}^{2}$-$m_{13}^{2}$ plane as the laboratory rotates with the Earth. As the
laboratory rotates, the structure of the $3\times3$ matrix of anisotropy coefficients $c_{jk}$ will
change. However, to analyze
this will necessarily mean consideration of more complicated $c_{jk}$ textures than we have so far
considered here; even if the form of Lorentz violation may be described by a single parameter $c$ at one
particular moment, at a different sidereal time, the rotation of the laboratory will change the
structure of the lab-frame $c_{jk}$ coefficient matrix. There have been some previous consideration
of how reaction thresholds for particle processes would change with the Earth's rotation, but the Dalitz
plot provides a more general structure. It would be possible, for instance, to look at the $m_{23}^{2}$
threshold for a decay---which is represented by the left-most point on the outline of a Dalitz plot---and
how this
quantity depends on a given type of Lorentz violation. However, the full outline gives significantly more
information than just observation of a single threshold.

In figures~\ref{fig-Dalitz2} and~\ref{fig-Dalitz3}, we show the analogous shape of the Dalitz plots
for hypothetical higher-energy decays. Figure~\ref{fig-Dalitz2} shows a moderately more relativistic
case, with $\sqrt{s}/m=6.5$, while figure~\ref{fig-Dalitz3} displays a Dalitz plot for a situation
in which the daughter pions are in the strongly relativistic regime. As the shape becomes increasingly
triangular, we can see that it correspondingly becomes relatively less sensitive to the variety of Lorentz
violation that we have considered. This appears to be a generic feature of $c_{\mu\nu}$-type Lorentz
violation. This is perhaps not unexpected, with a $c_{\mu\nu}$ coefficient tending to shift the boundaries
of kinematically allowed regions by less as the pion masses become less important and the energy-momentum
relations for the outgoing particles become more nearly linear.

\section{Conclusions and Outlook}
\label{sec-concl}

Dalitz plots are widely used in the analysis of three-body decays (especially decays into spinless
particles). The histogram in the interior of the Dalitz plot region is sensitive to variations
(particular resonance thresholds) in
the decay matrix element, while the outline of that region is determined by the decay kinematics. We
have looked at how this outline is affected by Lorentz violation of the $c_{\mu\nu}$ type, focusing
on specific illustrative models with broken boost and rotation symmetries. The Lorentz violation
changes the shape of the outline for three-pion decays, with more pronounced fractional changes
at lower values of $\sqrt{s}/m$.

Lorentz violation affecting unstable particle species is much harder to bound than similar violations
for the constituents of everyday matter. The shapes of Dalitz plots for decays into heavy mesons may
be good observables for constraining SME coefficients in those sectors. For realistic studies, it could be
useful to construct separate Dalitz plots in bins of sidereal time, so that evidence of changes in the
shape as a source beam's direction changes. It would also be natural to consider more
general textures for the tensor background $c_{\mu\nu}$.

Future analyses of three-body decays involving Lorentz violation may focus particularly on muon decays,
$\mu^{\pm}\rightarrow e^{\pm}+\nu+\bar{\nu}$. Muon physics has proved to be a fruitful source for
bounds on coefficients in sectors of the SME that are otherwise hard to constrain.
In~\cite{ref-noordmans3}, the authors used muon decay as a tool to test the invariance of the weak
interaction under Lorentz transformations. Further motivation
for investigating muon properties comes from the phenomenology of the muon anomalous magnetic moment
(``$g-2$'')~\cite{ref-aguillard,ref-colangelo1} and muonic hydrogen~\cite{ref-antognini},
where other puzzling deviations from predictions based on the standard model presently exist. Especially
for the magnetic moment experiment, where the direct observable is the decay rate, a full understanding
of the three-body decay is important.

The muon work was part of a more general development of the use of three-body
$\beta$-decays to explore Lorentz violations in the weak
sector~\cite{ref-noordmans,ref-noordmans4,ref-vos1,ref-vos2,ref-vos3,ref-sytema}, with the SME coefficients
only appearing in the virtual $W^{\pm}$ propagator, rather than the external particle kinematics.
The kinematic effects in weak three-body decays have also been studied~\cite{ref-diaz1}; however, most
kinematic analyses of Lorentz violation in $\beta$-decays have focused on two-body decays, such as
$\pi^{+}\rightarrow \mu^{+}+\nu$; or on Cerenkov-like emission of neutrino-antineutrino pairs, as in the
normally forbidden process $p^{+}\rightarrow p^{+}+\nu+\bar{\nu}$, for which the kinematics are
(because the neutrinos are nearly massless) essentially the same as those for a Cerenkov process.

Analyses of the effects of Lorentz violation on the Dalitz plots for weak decays are more complicated
than what we have considered in this paper, for a couple of different reasons. Firstly, the daughter
particles in such decays are not identical, so the SME coefficients for multiple sectors of the standard
model will come into play. Secondly, the decays involve fermions, for which there are additional
spin-dependent SME parameters beyond the $c_{\mu\nu}$.
Work on understanding the effects of these complexities is ongoing.

\section*{Acknowledgments}

The authors are grateful to M. Schindler for his helpful suggestions.


\begin{thebibliography}{99}

\bibitem{ref-kost18}V. A. Kosteleck\'{y}, S. Samuel, Phys. Rev. D {\bf 39}, 683
(1989).
\bibitem{ref-kost19}V. A. Kosteleck\'{y}, R. Potting, Nucl. Phys. B {\bf 359}, 545
(1991).
\bibitem{ref-gambini}R. Gambini, J. Pullin, Phys. Rev. D {\bf 59}, 124021 (1999).
\bibitem{ref-klinhamer3}F. R. Klinkhamer, Nucl. Phys. B \textbf{535}, 233 (1998).
\bibitem{ref-mocioiu}I. Mocioiu, M. Pospelov, R. Roiban, Phys. Lett. B {\bf 489},
390 (2000).
\bibitem{ref-amelino22}G. Amelino-Camelia, S. Majid, Int. J. Mod. Phys. A \textbf{15}, 4301 (2000).
\bibitem{ref-klinkhamer4}F. R. Klinkhamer, Nucl. Phys. B \textbf{578}, 277 (2000).
\bibitem{ref-carroll3}S. M. Carroll, J. A. Harvey, V. A. Kosteleck\'{y}, C. D.
Lane, T. Okamoto, Phys. Rev. Lett. {\bf 87}, 141601 (2001).
\bibitem{ref-alfaro}J. Alfaro, H. A. Morales-T\'{e}cotl, L. F. Urrutia, Phys. Rev.
D {\bf 65}, 103509 (2002).
\bibitem{ref-klinkhamer}F. R. Klinkhamer, C. Rupp, Phys. Rev. D {\bf 70}, 045020 (2004).
\bibitem{ref-bojowald}M. Bojowald, H. A. Morales-T\'{e}cotl, H. Sahlmann, Phys. Rev. D {\bf 71}, 084012 (2005).
\bibitem{ref-bernadotte}S. Bernadotte, F. R. Klinkhamer, Phys. Rev. D \textbf{75}, 024028 (2007).
\bibitem{ref-hossenfelder}S. Hossenfelder, Adv. High Energy Phys. \textbf{2014}, 950672 (2014).
\bibitem{ref-ghosh}K. J. B. Ghosh, F. R. Klinkhamer, Nucl. Phys. B \textbf{926}, 335 (2018).
\bibitem{ref-kost1}D. Colladay, V. A. Kosteleck\'{y}, Phys. Rev. D {\bf 55},
6760 (1997).
\bibitem{ref-kost2}D. Colladay, V. A. Kosteleck\'{y}, Phys. Rev. D {\bf 58},
116002 (1998).
\bibitem{ref-kost12}V. A. Kosteleck\'{y}, Phys. Rev. D, {\bf 69} 105009 (2004).
\bibitem{ref-jackiw5}R. Jackiw, S.-Y. Pi, Phys. Rev. D {\bf 68}, 104012 (2003).
\bibitem{ref-guarrera}D. Guarrera, A. J. Hariton, Phys. Rev. D {\bf 76}, 044011 (2007).
\bibitem{ref-alexander1}S. Alexander, N. Yunes, Phys. Rev. Lett. {\bf 99}, 241101 (2007).
\bibitem{ref-greenberg}O. W. Greenberg, Phys. Rev. Lett. {\bf 89}, 231602 (2002).
\bibitem{ref-schreck7}M. Shreck, J. A. A. S. Reis, Phy. Rev D \textbf{103}, 095029 (2021).
\bibitem{ref-schwinger3}J. S. Schwinger, Phys. Rev. \textbf{125} (1962) 397.
\bibitem{ref-callan1}C. G. Callan, R. F. Dashen, D. J. Gross, Phys. Lett. B \textbf{63} (1976), 334. 
\bibitem{ref-coleman}S. Coleman, S. L. Glashow, Phys. Rev. D {\bf 59}, 116008 (1999).
\bibitem{ref-jackiw1}R. Jackiw, V. A. Kosteleck\'{y}, Phys. Rev. Lett. {\bf 82}, 3572 (1999).
\bibitem{ref-chung1}J. M. Chung, Phys. Lett. B {\bf 461}, 138 (1999).
\bibitem{ref-victoria2}M. P\'{e}rez-Victoria, JHEP {\bf 04}, 032 (2001).
\bibitem{ref-kost4}V. A. Kosteleck\'{y}, C. D. Lane, A. G. M. Pickering,
Phys. Rev. D {\bf 65}, 056006 (2002).
\bibitem{ref-altschul2-PV}B. Altschul, Phys. Rev. D {\bf 70}, 101701 (R) (2004).
\bibitem{ref-altschul37}B. Altschul, Phys. Rev. D {\bf 99}, 125009 (2019).

\bibitem{ref-alt+oco}J. O'Connor, B. Altschul Phys. Rev. D \textbf{109}, 045005 (2024). 
\bibitem{ref-altschul4}B. Altschul, D. Colladay, Phys. Rev. D {\bf 71}, 125015 (2005).
\bibitem{ref-kost5}D. Colladay, V. A. Kosteleck\'{y}, Phys. Lett. B {\bf 511} 209 (2001).
\bibitem{ref-altschul2}B. Altschul, Phys. Rev. D \textbf{70}, 056005 (2004).
\bibitem{ref-lehnert2}R. Lehnert, R. Potting, Phys. Rev. D {\bf 70}, 125010 (2004);
erratum {\em ibid.} {\bf 70}, 129906 (2004).
\bibitem{ref-altschul12}B. Altschul, Phys. Rev. D {\bf 75}, 105003 (2007).
\bibitem{ref-schreck4}M. Schreck, Phys. Rev D \textbf{96}, 095026 (2017).
\bibitem{ref-bluhm4}R. Bluhm, V. A. Kosteleck\'{y}, C. D. Lane, N. Russell, Phys.
Rev. D {\bf 68}, 125008 (2003).
\bibitem{ref-ditsche}C. Ditsche, B. Kubis, Ulf-G. Mei{\ss}er, Eur. Phys. J. C \textbf{60}, 83 (2009).
\bibitem{ref-zeta}A. Alonso Izquierdo, W. Garcia Fuertes, M. A. Gonzalez Leon, J. Mateos Guilarte,
Nucl. Phys. B \textbf{635}, 525 (2002).
\bibitem{ref-kamand1}R. Kamand, B. Altschul, M. R. Schindler, Phys. Rev. D {\bf 95}, 056005 (2017).
\bibitem{ref-abt}I. Abt, et al. (ZEUS Collaboration), Phys. Rev. D \textbf{107}, 092008 (2023).
\bibitem{ref-lunghi1}E. Lunghi, N. Sherrill, Phys. Lett. B \textbf{862}, 139366 (2025).
\bibitem{ref-tables}V. A. Kosteleck\'{y}, N. Russell, Rev. Mod. Phys. {\bf 83}, 11 (2011);
updated as arXiv:0801.0287v18.
\bibitem{ref-anisovich}A. V. Anisovich, H. Leutwyler, Phys. Lett. B \textbf{375}, 335 (1996).
\bibitem{ref-bijnens}J. Bijnens, K. Ghorbani, JHEP \textbf{11}, 030 (2007).
\bibitem{ref-kampf1}Karol Kampf, Marc Knecht, Ji\v{r}\'{i} Novotn\'{y}, Martin Zdr\'{a}hal,
Phys. Rev. D \textbf{84}, 114015 (2011).
\bibitem{ref-colangelo2}G. Colangelo, S. Lanz, H. Leutwyler, E. Passemar, Phys. Rev. Lett. \textbf{118},
022001 (2017).
\bibitem{ref-ablikim}M. Ablikim, et al. (BESIII Collaboration), Phys. Rev. D \textbf{107}, 092007 (2023).

\bibitem{ref-noordmans3}J. P. Noodmans, C. J. G. Onderwater, H. W. Wilschut, R. G. E. Timmermans,
Phys. Rev. D \textbf{93}, 116001 (2016).
\bibitem{ref-aguillard}D. P. Aguillard, et al. (Muon $g-2$ Collaboration), Phys. Rev. D \textbf{110},
032009 (2024).
\bibitem{ref-colangelo1}G. Colangelo, et al., arXiv:2203.15810.
\bibitem{ref-antognini} A. Antognini, et al., Science \textbf{339}, 417 (2013).
\bibitem{ref-noordmans}J. P. Noordmans, H. W. Wilschut, R. G. E. Timmermans, Phys. Rev. C \textbf{87}, 055502
(2013).
\bibitem{ref-noordmans4}J. P. Noordmans, H. W. Wilschut, R. G. E. Timmermans, Phys. Rev. Lett.
\textbf{111}, 171601 (2013).
\bibitem{ref-vos1}K. K. Vos, H. W. Wilschut, R. G. E. Timmermans, Phys. Rev. C \textbf{91}, 038501 (2015).
\bibitem{ref-vos2}K. K. Vos, H. W. Wilschut, R. G. E. Timmermans, Phys. Rev. C \textbf{92}, 052501 (R) (2015).
\bibitem{ref-vos3}K. K. Vos, H. W. Wilschut, R. G. E. Timmermans, Rev. Mod. Phys. \textbf{87}, 1483 (2015).
\bibitem{ref-sytema}A. Sytema, et al., Phys. Rev. C \textbf{94}, 025503 (2016).
\bibitem{ref-diaz1}J. S. D\'{i}az, Adv. High Energy Phys. \textbf{2014}, 305298 (2014).






\end{thebibliography}
\end{document}